\documentclass[aps,prd,preprint,superscriptaddress,nofootinbib]{revtex4-1}
\pdfoutput=1
\usepackage{graphicx}  
\usepackage{latexsym}
\usepackage{slashed}
\usepackage{dcolumn}   
\usepackage{bm}        
\usepackage{amsmath}
\usepackage{amssymb}   
\usepackage{longtable}
\usepackage{float}

\newcommand{\lc}{\affiliation{Department of Chemistry and Physics, LaGrange College, LaGrange, GA 30240, USA}}

\begin{document}

\title{The Viability of Phantom Dark Energy as a Quantum Field in 1st-Order FLRW Space}

\author{Kevin J. Ludwick}
\email{kludwick@lagrange.edu} \lc

\begin{abstract}

In the standard cosmological framework of the 0th-order FLRW metric and the use of perfect fluids in the stress-energy tensor, 
dark energy with an equation-of-state parameter $w < -1$ (known as phantom dark energy)
implies negative kinetic energy and vacuum instability when modeled as a scalar field.  
However, the accepted values for present-day $w$ from Planck and WMAP9 include a significant range of values less than $-1$.  A flip of the sign in front of the kinetic 
energy term in the Lagrangian remedies the negative kinetic energy but introduces ghostlike instabilities, which perhaps may be rendered unobservable, but certainly not 
without great cost to the theory.  Staying within the confines of observational constraints 
and general relativity, for which there is good experimental validation, we consider 
a reasonable departure from the standard 0th-order framework in an attempt to see if negative kinetic energy can be avoided despite an apparent $w<-1$,
all without flipping the sign of the kinetic energy term.  
We consider a more accurate description of the universe through the perturbing of the isotropic and homogeneous FLRW metric and the components 
of the stress-energy tensor to 1st order.  We treat dark energy as a quantum scalar field in the background of this 1st-order
FLRW space-time, find an approximation for the Green's function, and calculate the expectation
value of the field's kinetic energy for $w<-1$
using adiabatic expansion to renormalize and obtain a finite value.  We find that the kinetic energy is positive for values of $w$ less than $-1$ in 0th- or 1st-order FLRW space, thus giving more theoretical credence to observational 
values of $w<-1$ and demonstrating that phantom dark energy does not categorically have negative kinetic energy.  For a non-minimal coupling parameter $\xi=0$, kinetic energy is positive for $w \gtrsim -1.22$, which includes virtually all values of constant $w$ allowed by cosmological data constraints, and more negative values of $w$ give positive kinetic energy for non-zero values of $\xi$.  Also, our results are generally applicable for a massive free field or a field with a small potential in a 0th- or 1st-order FLRW background dominated by a fluid with a constant $w$.    


\end{abstract}

\maketitle

\renewcommand{\thepage}{\arabic{page}}
\setcounter{page}{1}


\begin{center}
{\bf Introduction}
\end{center}

A recent milestone in observational cosmology happened when the High-z Supernova Search Team in 1998 \cite{Riess} and the Supernova Cosmology Project in 1999 \cite{SCP} published observations of the emission spectra of Type Ia supernovae indicating that the universe's rate of outward expansion is increasing.  
Galaxy surveys and the late-time integrated Sachs-Wolfe effect also give evidence for the universe's acceleration.  Thus, "dark energy" was proposed 
as the pervasive energy in the universe necessary to produce the outward force that causes this acceleration, which has been observationally tested and vetted 
since its discovery.  The 2011 Nobel Prize in Physics was awarded to Schmidt, Riess, and Perlmutter for their pioneering work leading to the discovery of dark energy.  
The present-day equation-of-state parameter $w$ from the equation of state most frequently tested by cosmological probes, $p=w \rho$ with constant $w$, assuming a 
flat universe and a perfect fluid representing dark energy, has been constrained by Planck in early 2015 to be $w=-1.006 \pm 0.045$ \cite{Planck}. 
The value from the Nine-Year Wilkinson Microwave Anisotropy Probe (WMAP9), combining data from 
WMAP, the cosmic microwave background (CMB), baryonic acoustic oscillations (BAO), supernova measurements, and $H_0$ measurements, is 
$w = -1.084 \pm 0.063$ \cite{WMAP9}.  Type Ia supernovae data at low redshift set more stringent constraints than CMB data on $w$ since dark energy dominates the dynamics of the universe at late times, as can be seen from these and other reported values.  In any case, we see that it is possible that $w<-1$.  

However, dark energy modeled as a perfect fluid with $w < -1$ leads to a field theory with negative kinetic energy (a ghost field theory), which implies vacuum instability.  Either 
the phantom ghost has positive density and violates unitarity, rendering it unphysical, or unitarity is satisfied and the density is negative, which leads to vacuum 
instability \cite{Cline}.  This phantom dark energy with a wrong-sign kinetic term described as an 
effective field theory may be able to make this instability unobservable, but not without great 
difficulty and perhaps sacrifice of well-accepted physical principles \cite{Cline, CHT}.  

One deduces the ghost nature of phantom dark energy from $w<-1$ within the standard cosmological framework of the 0th-order Friedmann-Lema\^{i}tre-Robertson-Walker 
(FLRW) metric with the use of perfect fluids 
in the stress-energy tensor, but the condition for negative kinetic energy is different for different frameworks.  Given that our universe is not 
perfectly isotropic and homogeneous, we examine the possibility for positive kinetic energy with $w<-1$ in light of first-order perturbations to 
the FLRW metric
 and the components of the stress-energy tensor.  In earlier work \cite{1507.06492}, we found that for certain classical scalar field models of dark energy
  with non-constant $w<-1$, it is possible to have positive 
 kinetic energy for certain length and time scales.  We also found that it was not possible to have positive kinetic energy for constant $w<-1$ because that condition implied that 
 one of the perturbations of the stress-energy tensor would have to be bigger than $1$, violating the assumption of perturbation.  In this work, we treat dark energy as a quantum scalar field with constant $w<-1$ and calculate the expectation value of the kinetic energy using 
the method of adiabatic subtraction for renormalization, and we find that the kinetic energy is positive for all relevant length and time scales.

 \bigskip
\begin{center}
{\bf Dark Energy as a Scalar Field}
\end{center}

We consider the Einstein-Hilbert action for general relativity with a real scalar field for dark energy ($c=1$):
\begin{equation}
\label{action}
S = \int d^4 x \sqrt{-g} \left[ \frac{R}{16 \pi G} - \frac{1}{2} g^{\mu \nu} \nabla_\mu \phi \nabla_\nu \phi -\frac{1}{2} m^2 \phi^2 -\frac{1}{2} \xi R \phi^2 - V(\phi) \right] + S_m,
\end{equation}
where the first term is the usual contribution to the Einstein tensor, the second term is the kinetic energy term, the third term is the mass term, the fourth term is the non-minimal coupling term (usually included 
for its utility in the renormalization of a scalar field in a curved background), the fifth term is 
the dark energy potential, and $S_m$ is the action 
for the rest of the components of the stress-energy tensor $T_{\mu \nu}$.  We would like to calculate the expectation value of the kinetic energy, which is, for a free-field theory,  
\begin{equation}
\label{expectation}
- \frac{1}{2} g^{\sigma \rho} <\partial_\rho \phi \partial_\sigma \phi > = - \frac{1}{2} g^{\sigma \rho} \partial_\rho  \partial_\sigma' i G(x,x')|_{x' \rightarrow x}, 
\end{equation} 
and we will renormalize via adiabatic subtraction.  To obtain the Green's function $G(x,x')$, we must solve the equation of motion \cite{0803.3684}
\begin{equation}
\label{equationofmotion}
(-\Box_x + m^2 +\xi R^2) G(x,x') = g^{-1/4}(x) \delta(x-x') g^{-1/4}(x'), 
\end{equation}
where the operator $\Box_x = g(x)^{-1/2} \frac{\partial}{\partial x^\mu} \left[ g(x)^{1/2} g(x)^{\mu \nu} \frac{\partial}{\partial x^\nu} \right]$ when applied to a scalar.  The data that 
support a value of $w<-1$ are for late-time redshift values for which dark energy dominates, as it does currently, so we consider dark energy to be the only component of the universe, which therefore specifies the background space.  
The self-interaction terms in $V(\phi)$ specify the equation of state of dark energy, and for constant $w$ satisfying $|w+1| \ll 1$, it can be 
shown that the potential is slowly varying and small in 0th-order FLRW space, and it should also be small in 1st-order FLRW space.  
So we will ignore the terms from $V(\phi)$ in our calculation 
of the kinetic energy, so we will simply evaluate Eq. (\ref{expectation}) to find the renormalized kinetic energy.

\bigskip
\begin{center}
{\bf The 1st-Order FLRW Metric}
\end{center}

We take the 1st-order flat FLRW metric with scalar perturbations in the synchronous gauge to be our fixed background (using the notation of \cite{BertMa}),
\begin{equation}
\label{metric}
ds^2 = a^2(\tau) \biggl[ -d\tau^2 + (\delta_{ij}+ h_{ij}) dx^i dx^j \biggr],
\end{equation}
with the scalar mode of $h_{ij}$ written in $k$-space as 
\begin{equation}
\label{metricperturb}
h_{ij}(\vec{x},\tau) = \int d^3k e^{i \vec{k} \cdot \vec{x}} \biggl\{\hat{k_i} \hat{k_j} h(\vec{k},\tau)+(\hat{k_i} \hat{k_j} -\frac{\delta_{ij}}{3}) 6 \eta(\vec{k},\tau) \biggr\}.
\end{equation}

The Friedmann equations resulting from solving Einstein's equation for the flat FLRW metric (0th order) are 
\begin{align}
\mathcal{H}^2~ &= ~\frac{8 \pi G}{3} a^2 \rho, \label{Feq1} \\
\dot{\mathcal{H}}~ &= ~ - \frac{4 \pi G}{3} a^2 (\rho+ 3P), \label{Feq2} 
\end{align}
where $\cdot$ denotes differentiation with respect to $\tau$, $\mathcal{H} \equiv \frac{\dot{a}}{a}$, and $\rho$ and $P$ represent the total
 density and pressure respectively.  
These equations lead to the evolution equation for 
each density component:
\begin{equation}
\label{densityevo}
\dot{\rho} = -3 \mathcal{H} (\rho+ P).
\end{equation}

The equations resulting from solving the perturbed Einstein equation in 
$k$-space to first order are
\begin{subequations}
\label{perturb}
\begin{align}
k^2 \eta - \frac{1}{2} \mathcal{H} \dot{h}~ &= ~ 4 \pi G a^2 \delta T^0_0, \label{perturb1} \\ 
k^2 \dot{\eta} ~ &= ~ 4 \pi G a^2 (\rho+P) \theta, \label{perturb2} \\
\ddot{h} + 2 \mathcal{H} \dot{h} - 2 k^2 \eta ~ &= ~ -8 \pi G a^2 \delta T^i_i, \label{perturb3} \\
\ddot{h} + 6 \ddot{\eta} +2 \mathcal{H} (\dot{h} + 6 \dot{\eta}) - 2k^2 \eta ~ &=~ -24 \pi G a^2 (\rho+P) \sigma, \label{perturb4}
\end{align}
\end{subequations}
where $\theta$ is the divergence of the fluid velocity 
$v_i$, $(\rho + P) \sigma \equiv - (\hat{k}_i \hat{k}_j - \frac{1}{3} \delta_{ij}) \Sigma^i_{~j}$ where 
$\Sigma^i_{~j}$ is the anisotropic shear perturbation, and $h$ and $\eta$ are the scalar modes of the metric perturbation.  The stress-energy 
tensor for a perturbed perfect fluid is given by 
\begin{align}
T^0_{~0} ~ &= ~ -(\rho+\delta \rho),  \nonumber \\
T^0_{~i} ~ &=~ (\rho+P)v_i,  \nonumber \\
T^i_{~j} ~&=~ (\rho+ \delta P)\delta^i_{~j} + \Sigma^i_{~j}, \quad \Sigma^i_{~i} = 0. \label{stressenergy}
\end{align}

The conservation of energy-momentum, $T^{\mu \nu}_{~~;\mu} = 0$, gives (using $\delta \equiv \delta \rho/\rho$)
\begin{subequations}
\label{cons}
\begin{align}
\dot{\delta} ~&= ~ -(1+w) \left(\theta + \frac{\dot{h}}{2} \right)- 3 \mathcal{H} \left( \frac{\delta P}{\delta \rho} -w \right) \delta, \label{perturb5} \\
\dot{\theta} ~&= ~ -\mathcal{H}(1-3w)\theta - \frac{\dot{w}}{1+w} \theta + \frac{\delta P/\delta \rho}{1+w} k^2 \delta - k^2 \sigma. \label{perturb6}
\end{align}
\end{subequations}
Eq. (\ref{cons}) is valid when considering each fluid component or the total fluid, but Eq. (\ref{perturb}) is valid only 
for the total fluid.  The anisotropic shear stress is $0$ ($\sigma = 0$) throughout, and in what follows, we use $c=G=1$.  

$\delta P/\rho$ for a given fluid component
is in general given by  
\begin{equation}
\label{deltap}
\frac{\delta P}{\rho} = c_s^2 \delta + (c_s^2-c_a^2) 3 \mathcal{H}(1+w) \frac{\theta}{k^2},
\end{equation}
where $c_s$ is the fluid's sound speed and $c_a^2 \equiv \dot{P}/\dot{\rho} = w + \dot{w} \rho/\dot{\rho}$ is defined as the square of the fluid's 
adiabatic sound speed \cite{0804.0232}.  For a barotropic fluid, $c_s^2 = c_a^2$, and $c_a^2 = w$ for constant $w$.  Even though dark energy can have a barotropic 
equation of state, 
treating it like an adiabatic fluid (for which Eq. (\ref{deltap}) reduces to $\delta P = c_s^2 \delta \rho$) would imply imaginary sound speed and instabilities in dark energy, 
so we use this general relation between $\delta P$ and $\delta \rho$.  

Since the dynamics of the universe is currently dominated by dark energy, we will assume dark energy is the only fluid component with an equation of state $P = w \rho$ where 
$w<-1$ and is a constant.  Of course, this assumption is an approximation since dark energy does not comprise 100\% of the universe's contents, so this assumption would technically only be valid in the asymptotic future time limit.  However, dark energy comprises a large percentage of our universe for recent times (currently about 68\%, with the percentage growing when $w<-1$), and one advantage of assuming only the presence of dark energy for recent and future times is that we can obtain analytic solutions for the metric perturbations that are valid for all length scales.  With this assumption, $\mathcal{H}$ can be written as 
\begin{equation}
\mathcal{H} = \frac{2}{(3w +1) \tau}.
\end{equation}
For $w < -1/3$, $\tau \in (- \infty,0)$ for $a \in (0, \infty)$.  We find it more convenient 
to work in terms of $a$ rather than $\tau$.  $\tau$ can be related to $a$ via Eq. (\ref{Feq1}).   
Using Eqs. (\ref{perturb1}), (\ref{perturb2}), (\ref{perturb5}), and (\ref{perturb6}), and keeping the relevant growing modes of perturbations, one finds that \cite{1507.06492} 
\begin{equation}
\label{h}
h(\vec{k},\tau)=2^{\frac{3(1+w)}{2(1+3w)}} S \left(\frac{\rho_{DE0} \pi}{3} \right)^{\frac{1-3w}{2+6w}} 4 \pi \rho_{DE0} k^{ - \frac{2}{3w+1} -1} a^{ - \frac{3(w+1)}{2}}
\end{equation}
and $\eta(\vec{k},\tau)=0$, where $\rho_{DE0} = \frac{3 H_0^2}{8 \pi} \Omega_{DE0}$ is the present-day density of dark energy and $S$ is a dimensionless constant of integration.  As a scaling factor 
for the perturbation magnitude, and from observational constraints from the primordial power spectrum, $|S| \ll 1$.

We can then use Eq. (\ref{metricperturb}) to arrive at the metric perturbations in spatial coordinates of the metric.  Using $r=\sqrt{x^2+y^2+z^2}$, the result is 
\begin{equation}
\label{hij_x}
h_{ij}(\vec{x},\tau) = \partial_i \partial_j \left[ 2^{\frac{3(1+w)}{2(1+3w)}} S \left(\frac{\rho_{DE0} \pi}{3} \right)^{\frac{1-3w}{2+6w}} 8 \pi^2 \rho_{DE0} a^{ - \frac{3(w+1)}{2}} \frac{(-i r)^{\frac{2}{1+3w}}+(ir)^{\frac{2}{1+3w}}}{r^2} \Gamma(1-\frac{2}{1+3w}) \right]
\end{equation}
where we have ignored the oscillatory upper bound of the integral with respect to $k$, and this is valid especially since we are only interested in large scales over which dark energy is relevant; dark energy is only observationally active on large scales during the epoch of dark energy domination, which is what we are considering here.    

\bigskip
\begin{center}
{\bf The Green's Function in Riemann Normal Coordinates}
\end{center}

One can see that the equation of motion, Eq. (\ref{equationofmotion}), cannot be solved exactly for our 1st-order FLRW metric, so we will express the equation of motion as a series 
expansion using Riemann normal coordinates \cite{ParkerToms}.  Assuming for any point $P$ in the neighborhood of $Q$ that there is a unique geodesic joining these points, we can use Riemann normal coordinates of that point $P$:  $y^\mu = \lambda \xi^\mu$, where $\xi^\mu$ is the tangent to the geodesic at the point $Q$, and $\lambda$ is a parameter representing how far off along the geodesic we are from $Q$.  We take the origin $Q$ at the space-time point $x'$, where $y^\mu=0$, and we denote $x^\mu$ to be its normal coordinate $y^\mu$.  We can therefore write $g_{\mu \nu}(y=0) = \eta_{\mu \nu}$, and its determinant is $|g(y=0)|=1$.  We expand $g_{\mu \nu}(y)$, $R(y)$, and $g(y)$ about $y^\mu=0$.  And the limit we will take in solving for the expectation value of the kinetic energy in Eq. (\ref{expectation}) $x' \rightarrow x$ is equivalent to $y \rightarrow 0$.  
Eq. (3.180) of \cite{ParkerToms} gives a series expansion of a tensor field about the origin in Riemann normal coordinates:
\begin{equation}
\label{tensorexpansion}
\begin{split}
&W_{\alpha_1 \cdots \alpha_p}(y) = W_{\alpha_1 \cdots \alpha_p}(0)+W_{\alpha_1 \cdots \alpha_p ; \mu}(0) y^\mu \\
&+\frac{1}{2!} \bigg[ W_{\alpha_1 \cdots \alpha_p ; \mu \omega} - \frac{1}{3} 
\sum\limits_{k=1}^{p} R^{\nu}_{\mu \alpha_k \omega} W_{\alpha_1 \cdots \alpha_{k-1} \nu \alpha_{k+1} \cdots \alpha_p} \bigg]_0 y^\mu y^\omega \\
&+ \frac{1}{3!} \bigg[ W_{\alpha_1 \cdots \alpha_p; \mu \omega \sigma} - \sum\limits_{k=1}^{p} R^{\nu}_{\mu \alpha_k \omega} W_{\alpha_1 \cdots \alpha_{k-1} \nu \alpha_{k+1} \cdots \alpha_p; \sigma} - \frac{1}{2} \sum\limits_{k=1}^{p} R^{\nu}_{\mu \alpha_k \omega; \sigma} W_{\alpha_1 \cdots \alpha_{k-1} \nu \alpha_{k+1} \cdots \alpha_p} \bigg]_0 y^\mu y^\omega y^\sigma + ...~,
\end{split}
\end{equation}
where each coefficient in front of the factors of $y$ are evaluated at $y=0$. 
From \cite{0001078}, the metric expansion about the origin in Riemann normal coordinates to 6th order is
\begin{equation}
\label{metricnormal}
\begin{split}
&g_{\alpha_1 \alpha_2}(y) = g_{\alpha_1 \alpha_2}(0) + \frac{1}{2!} \frac{2}{3}
R_{\alpha_1 \beta_1 \beta_2 \alpha_2}(0) y^{\beta_1} y^{\beta_2}  \\
&+ \frac{1}{3!} \Big[ \nabla_{\beta_1} R_{\alpha_1 \beta_2 \beta_3 \alpha_2} \Big]_0 y^{\beta_1} \cdots y^{\beta_3}  \\
&+ \frac{1}{4!}\frac{6}{5} \Big[\nabla_{\beta_1}\nabla_{\beta_2} R_{\alpha_1 \beta_3 \beta_4 \alpha_2} 
+ \frac{8}{9} R_{\alpha_1 \beta_1 \beta_2 \rho} R^{\rho}_{\beta_3 \beta_4 \alpha_2}\Big]_0 y^{\beta_1} \cdots y^{\beta_4} \\
&+ \frac{1}{5!} \frac{4}{3} \Big[\nabla_{\beta_1} \cdots \nabla_{\beta_3} R_{\alpha_1 \beta_4 \beta_5 \alpha_2} 
+ 2 \left(\nabla_{\beta_1} R_{\alpha_1 \beta_2 \beta_3 \rho} R^{\rho}_{\beta_4 \beta_5 \alpha_2} +  \alpha_1 \leftrightarrow \alpha_2 \right) 
\Big]_0 y^{\beta_1} \cdots y^{\beta_5}  \\
&+ \frac{1}{6!} \frac{10}{7}\Big[ \nabla_{\beta_1} \cdots \nabla_{\beta_4} R_{\alpha_1 \beta_5 \beta_6 \alpha_2} + \frac{17}{5}\left(\nabla_{\beta_1}\nabla_{\beta_2} 
R_{\alpha_1 \beta_3 \beta_4 \rho} 
R^{\rho}_{\beta_5 \beta_6 \alpha_2} + \alpha_1 \leftrightarrow \alpha_2 \right)  \\
&+ \frac{11}{2}\nabla_{\beta_1} R_{\alpha_1 \beta_2 \beta_3 \rho} \nabla_{\beta_4} R^{\rho}_{\beta_5 \beta_6 \alpha_2} +
\frac{8}{5}R_{\alpha_1 \beta_1 \beta_2 \rho} R^{\rho}_{\beta_3 \beta_4 \mu} R^{\mu}_{\beta_5 \beta_6 \alpha_2} \Big]_0 
y^{\beta_1} \cdots y^{\beta_6} + ...~ ,
\end{split}
\end{equation}
and the inverse metric is 
\begin{equation}
\label{inversemetricnormal}
\begin{split}
&g^{\alpha_1 \alpha_2}(y) =  g^{\alpha_1 \alpha_2}(0) - \frac{1}{2!} \frac{2}{3}
R^{\, \alpha_1 \quad \alpha_2}_{\,\,\,\,\,\,  \beta_1 \beta_2}(0) y^{\beta_1} y^{\beta_2}  \\
&- \frac{1}{3!} \Big[\nabla_{\beta_1} R^{\, \alpha_1 \quad \alpha_2}_{\,\,\,\,\,\,  \beta_2 \beta_3} \Big]_0 y^{\beta_1} \cdots y^{\beta_3}  \\
&- \frac{1}{4!}\frac{6}{5} \Big[\nabla_{\beta_1}\nabla_{\beta_2} R^{\, \alpha_1 \quad \alpha_2}_{\,\,\,\,\,\,  \beta_3 \beta_4} 
- \frac{4}{3} R^{\alpha_1}_{\, \beta_1 \beta_2 \rho} R^{\, \rho \quad \,\, \alpha_2}_{\,\,\,\,\, \beta_3 \beta_4}\Big]_0 
y^{\beta_1} \cdots y^{\beta_4} \\
&- \frac{1}{5!} \frac{4}{3} \Big[\nabla_{\beta_1} \cdots \nabla_{\beta_3} R^{\, \alpha_1 \quad \alpha_2}_{\,\,\,\,\,\, \beta_4 \beta_5} 
- 3 \left(\nabla_{\beta_1} R^{\alpha_1}_{\, \beta_2 \beta_3 \rho} R^{\, \rho \quad \,\, \alpha_2}_{\,\,\,\, \beta_4 \beta_5} +  
\alpha_1 \leftrightarrow \alpha_2 \right) \Big]_0 y^{\beta_1} \cdots y^{\beta_5}  \\
&- \frac{1}{6!} \frac{10}{7}\Big[ \nabla_{\beta_1} \cdots \nabla_{\beta_4} R^{\, \alpha_1 \quad \alpha_2}_{\,\,\,\,\,\, \beta_5 \beta_6} 
- 5\left(\nabla_{\beta_1}\nabla_{\beta_2} R^{\alpha_1}_{\, \beta_3 \beta_4 \rho} 
R^{\, \rho \quad \, \alpha_2}_{\,\,\,\, \beta_5 \beta_6} + \alpha_1 \leftrightarrow \alpha_2 \right)  \\
&- \frac{17}{2}\nabla_{\beta_1} R^{\alpha_1}_{\beta_2 \beta_3 \rho} \nabla_{\beta_4} R^{\rho \quad \,\, \alpha_2}_{\,\,\,\, \beta_5 \beta_6} +
\frac{16}{3}R^{\alpha_1}_{\beta_1 \beta_2 \rho} R^{\rho }_{\beta_3 \beta_4 \mu} 
R^{\mu \quad \alpha_2}_{\beta_5 \beta_6} \Big]_0 y^{\beta_1} \cdots y^{\beta_6} - ...~.
\end{split}
\end{equation}
Using Eq. (\ref{tensorexpansion}), it follows that 
\begin{equation}
\label{Rexpansion}
\begin{split}
&R(y)=R(0)+\Big[\nabla_{\beta_1}R \Big]_0 y^{\beta_1} +\frac{1}{2!} \Big[\nabla_{\beta_1} \nabla_{\beta_2}R \Big]_0 y^{\beta_1} y^{\beta_2}
+\frac{1}{3!} \Big[\nabla_{\beta_1} \cdots \nabla_{\beta_3} R \Big]_0 y^{\beta_1} \cdots y^{\beta_3} \\
&+\frac{1}{4!} \Big[\nabla_{\beta_1} \cdots \nabla_{\beta_4} R \Big]_0 y^{\beta_1} \cdots y^{\beta_4}
+\frac{1}{5!} \Big[\nabla_{\beta_1} \cdots \nabla_{\beta_5} R \Big]_0 y^{\beta_1} \cdots y^{\beta_5}
+\frac{1}{6!} \Big[\nabla_{\beta_1} \cdots \nabla_{\beta_6} R \Big]_0 y^{\beta_1} \cdots y^{\beta_6} + ...~ .
\end{split}
\end{equation}

To obtain a series expansion for the determinant $g(y)$, we can use the relation $\ln(\det X)=\mathrm{tr}(\ln X)$ for a matrix $X$ and the 
following relations for an invertible matrix $A$ and matrix $B$:  
\begin{equation}
\begin{split}
&\det(A+B)=e^{\ln \det(A+B)}=e^{\ln [\det A \det(I+A^{-1}B)]}=  e^{\ln \det A + \ln \det(I+A^{-1}B)} = \det A e^{\mathrm{tr} \ln(I+A^{-1}B)} \\
&\ln(I+A^{-1}B) = A^{-1}B - \frac{(A^{-1}B)^2}{2}+\frac{(A^{-1}B)^3}{3} - ... ~.
\end{split}
\end{equation}
Via Eq. (\ref{metricnormal}), $g_{\mu \nu}(y) = A + B$, where $A = g_{\mu \nu}(0) = \eta_{\mu \nu}$ and $B$ is the matrix represented by all the other terms in Eq. (\ref{metricnormal}).  Then using the expansion $e^x = 1+x+\frac{x^2}{2!}+\frac{x^3}{3!} + ...$, we can obtain $g(y)$.  It turns out we only need our expansions to 5th order (as will be explained later).  With 
the help of the {\it Mathematica} package xTensor \cite{0807.0824, 1110.2662}, we obtain
\begin{equation}
\begin{split}
\label{gdet}
&g(y) = 1 - \frac{1}{3} R_{\beta_1 \beta_2}(0) y^{\beta_1} y^{\beta_2} \\
&- \frac{1}{6} \Big[\nabla_{\beta_3} R_{\beta_1 \beta_2} \Big]_0  y^{\beta_1} \cdots y^{\beta_3} \\
&+ \frac{1}{360} \Big[ 20 R_{\beta_1 \beta_2} R_{\beta_3 \beta_4} +16 R_{\beta_1 \,\, \beta_2 \rho}^{\,\,\,\, \sigma} R_{\beta_3 \sigma \beta_4}^{\quad \,\,\,\,\,\, \rho} 
+18 \nabla_{\beta_3} \nabla_{\beta_4} R_{\beta_1 \beta_2} -20 R^{\rho}_{\,\,\,\, \beta_1 \sigma \beta_3} R^{\sigma}_{\,\,\,\, \beta_2 \rho \beta_4}  \Big]_0 y^{\beta_1} \cdots y^{\beta_4}\\
&+ \frac{1}{360} \Big[ 16 R^{\rho \quad \,\,\, \sigma}_{\,\,\,\, \beta_4 \beta_5} \nabla_{\beta_1} R_{\sigma \beta_2 \beta_3 \rho} -4 \nabla_{\beta_1} \nabla_{\beta_2} \nabla_{\beta_3} R_{\beta_4 \beta_5} +20 R_{\beta_1 \beta_2} \nabla_{\beta_5} R_{\beta_3 \beta_4} \\
&- 20 R^{\rho}_{\,\,\,\, \beta_1 \sigma \beta_3} \nabla_{\beta_5} R^{\sigma}_{\,\,\,\, \beta_2 \rho \beta_4}  \Big]_0  y^{\beta_1} \cdots y^{\beta_5} + ... ~.
\end{split}
\end{equation}

For mathematical simplification, as is done in \cite{ParkerToms}, we define 
\begin{equation}
G(x,x') = g(x)^{-1/4} \bar{G}(x,x') g(x')^{-1/4}
\label{Gbar}
\end{equation}
and make use of the generalized Fourier transformation 
\begin{equation}
\bar{G}(x,x') = \int \frac{d^nk}{(2 \pi)^n} e^{iky} \bar{G}(k),
\label{Gmom}
\end{equation}
where $ky = \eta^{\mu \nu} k_\mu y_\nu$.  Then we express $\bar{G}(k)$ as
\begin{equation}
\bar{G}(k) = \bar{G}_0 (k) + \bar{G}_1 (k) + \bar{G}_2 (k) + ...~,
\end{equation}
where $\bar{G}_i (k)$ involves $i$ derivatives of the metric.  

For a given interval $x$ to $x'$, our adiabatic assumption is that the rate of change of $a(t)$ is sufficiently slow, or adiabatic.  So each higher-order term in 
metric derivative should be smaller than the previous.  

Using these expansions about the origin, we may express Eq. (\ref{equationofmotion}) in momentum space and solve iteratively for $\bar{G}_i$ of $i$th adiabatic order.  We obtain 
\begin{equation}
\bar{G}_0(k) = (k^2 + m^2)^{-1}.
\end{equation}  
At 0th order, the Green's function is that of a free field in Minkowski space, as expected.  We obtain $\bar{G}_1(k) =0$.  In solving for higher-order $\bar{G}_i$, we use the fact that 
$\partial/ \partial y^\mu \rightarrow i k_\mu$ in momentum space, and $y^\mu \rightarrow i \partial/ \partial k_\mu$, which follows from integration by parts as the boundary term 
goes to 0 since $\bar{G}_i(k)$ vanishes for large $k^2$.  For 2nd adiabatic order, we obtain
\begin{equation}
\bar{G}_2(k)= \left( \frac{1}{6} - \xi \right) R(0) (k^2 +m^2)^{-2}.
\end{equation}
We see that $\bar{G}_i (k)$ is of order $k^{-(2+i)}$.  It turns out that we need solve iteratively at least up to adiabatic order $i=5$ in $4$ dimensions in order to subtract out all 
ultraviolet divergences and keep some finite part for the kinetic energy.  Using Eqs. (\ref{Gbar}) and (\ref{Gmom}) in Eq. (\ref{expectation}), the kinetic energy is 
\begin{equation}
\label{integralKE}
- \frac{1}{2} g^{\sigma \rho} \frac{\partial}{\partial x^{\rho}}  \frac{\partial}{\partial x'^{\sigma}} i G(x,x')|_{x' \rightarrow x} = - \frac{1}{2} g^{\sigma \rho}(y) \frac{\partial}{\partial x'^{\sigma}  } \int \frac{d^4 k}{(2 \pi)^4} e^{iky} (ik_\rho) i \bar{G}(k) \big|_{y \rightarrow 0},
\end{equation}
where $\frac{\partial}{\partial x^{\rho}} \rightarrow i k_{\rho}$ and $x'$ refers to the parts evaluated at $y=0$, and the $x'$-dependent parts of $\bar{G}(k)$ are the curvature factors that are evaluated at $y=0$.  The integral diverges for $i \leq 3$ and converges for $i=4$ or bigger, and since the integral of the term involving $\bar{G}_4$ gives a contribution of $0$, we must go to at least order $i=5$.  So using the method of adiabatic subtraction, we calculate a physically meaningful (renormalized) approximation 
of the kinetic energy $KE_{phys}$ as  
\begin{equation}
KE_{phys} = \sum KE_{all ~orders} - \sum KE_{divergent} \approx KE_{i=4 ~ and~ i=5},
\end{equation}
and $i>5$ contributions would be smaller than the $i=5$ contribution.  After solving iteratively up to $i=5$, we obtain (with the understanding that all the metric curvature factors in the following are evaluated at $y=0$)
\begin{equation}
\label{G5}
\begin{split}
&\bar{G}(k) = (k^2+m^2)^{-1} + \left( \frac{1}{6} - \xi \right) R (k^2+m^2)^{-2} +i C_\alpha (k^2 + m^2)^{-1} \frac{\partial}{\partial k_\alpha} (k^2+m^2)^{-1}  \\ 
&+  \left( \frac{1}{6} - \xi \right)^2 R^2 (k^2+m^2)^{-3} - E_{\alpha \beta} (k^2+m^2)^{-1} \frac{\partial}{\partial k_\alpha} \frac{\partial}{\partial k_\beta} (k^2+m^2)^{-1}   \\
&+i C_\alpha (k^2+m^2)^{-1} \bigg( \frac{1}{6} - \xi \bigg) R \frac{\partial}{\partial k_\alpha} (k^2+m^2)^{-2}+i \bigg( \frac{1}{6} - \xi \bigg) R C_\alpha (k^2+m^2)^{-2} \frac{\partial}{\partial k_\alpha} (k^2+m^2)^{-1}   \\ 
&+\frac{i}{3} R_{\alpha}^{\,\, \nu} (k^2+m^2)^{-1} k_\nu C_\beta \frac{\partial}{\partial k_\alpha} \bigg[ (k^2+m^2)^{-1} \frac{\partial}{\partial k_\beta} (k^2+m^2)^{-1} \bigg] \\
&+ \frac{i}{3} R^{\mu \,\, \nu}_{\,\, \alpha \,\, \beta} (k^2+m^2)^{-1} k_\mu k_\nu C_\gamma \frac{\partial}{\partial k_\alpha} \frac{\partial}{\partial k_\beta} \bigg[ (k^2+m^2)^{-1} \frac{\partial}{\partial k_\gamma} (k^2+m^2)^{-1} \bigg] \\
& - i I_{\alpha \beta \gamma} (k^2+m^2)^{-1} \frac{\partial}{\partial k_\alpha} \frac{\partial}{\partial k_\beta} \frac{\partial}{\partial k_\gamma} (k^2+m^2)^{-1},  
\end{split}
\end{equation}
where, letting the semicolon denote the application of the covariant derivative, we use the following:
\begin{equation}
\label{C}
\begin{split}
&C_\alpha \equiv \frac{1}{6} R_{\alpha \,\,\, ; \beta}^{\,\,\, \beta} +\frac{1}{12} R_{; \alpha} - \xi R_{; \alpha} \\
&E_{\alpha \beta} \equiv -\frac{1}{12} R_{\alpha \gamma} R_{\beta}^{\,\,\, \gamma} + \frac{1}{15} R^{\gamma \delta} R_{\alpha \gamma \beta \delta}+ \frac{1}{90} R_{\alpha}^{\,\,\, \gamma \delta \epsilon} R_{\beta \gamma \delta \epsilon} + (\frac{1}{40}- \frac{1}{2} \xi )R_{; \alpha \beta} + \frac{1}{20} R_{\alpha \,\,\, ; \beta \gamma}^{\,\,\, \gamma}  + \frac{1}{90} R_{\alpha}^{\,\,\, \gamma \delta \epsilon} R_{\beta \delta \gamma \epsilon} \\
&+ \frac{1}{20} R_{\alpha \,\,\, ; \gamma \beta}^{\,\,\, \gamma} +\frac{1}{40} \Box R_{\alpha \beta}  \\
&I_{\alpha \beta \gamma} \equiv \frac{1}{30} R_{\alpha \delta \beta \epsilon} R^{\delta \epsilon}_{\,\,\,\,\, ; \gamma} - \frac{1}{12} R_{\alpha \delta ; \beta} R^{\delta}_{\,\, \gamma} + \frac{1}{30} R_{\alpha \delta \beta \epsilon ; \gamma} R^{\delta \epsilon} + \frac{1}{90} R_{\alpha \delta \epsilon \kappa ; \beta} R_{\gamma}^{\,\,\, \delta \epsilon \kappa} \\ 
&+ \frac{1}{90} R_{\alpha \epsilon \delta \kappa ; \beta} R_{\gamma}^{\,\,\, \delta \epsilon \kappa} + (\frac{1}{180}  - \frac{1}{6} \xi) R_{; \alpha \beta \gamma} + \frac{1}{90} R_{\alpha \,\,\, ; \delta \beta \gamma}^{\,\,\, \delta} + \frac{1}{90} R_{\alpha \,\,\, ; \beta \delta \gamma}^{\,\,\, \delta} \\
&+ \frac{1}{180} (\Box R_{\alpha \beta})_{; \gamma} - \frac{1}{36} R_{\alpha}^{\,\,\, \delta}  R_{\beta \gamma ; \delta} + \frac{1}{90} R_{\alpha \,\,\, ; \beta \gamma \delta}^{\,\,\, \delta} + \frac{1}{36} R_{\alpha}^{\,\,\, \delta} R_{\beta \delta \gamma \,\, ; \epsilon}^{\,\,\,\,\,\,\,\,\,\, \epsilon} \\
&+ \frac{1}{180} \eta^{\delta \epsilon} R_{\alpha \beta ; \delta \gamma \epsilon} + \frac{1}{180} \Box(R_{\alpha \beta ; \gamma}) + \frac{1}{18} R_{\alpha \,\,\, ; }^{\,\,\, \delta \,\,\, \epsilon} R_{\beta \delta \gamma \epsilon} + \frac{1}{90} R_{\alpha}^{\,\,\, \delta \epsilon \kappa} R_{\beta \delta \gamma \epsilon; \kappa} - \frac{1}{90} R_{\alpha \,\,\, \beta}^{\,\,\, \delta \,\,\,\,\,\, \epsilon} R_{\gamma \delta \epsilon \,\,\, ; \kappa}^{\,\,\,\,\,\,\,\,\, \kappa}.
\end{split} 
\end{equation}

After applying the $x'$-derivative in Eq. (\ref{integralKE}) (keeping only the physically relevant $i=4$ and $i=5$ terms), we use integration by parts to express all the $k$-dependence in the integral as $(k^2+m^2)^{-3}$.  We then take the limit $y \rightarrow 0$ (which makes $e^{i k y} \rightarrow 1$ and $g^{\sigma \rho}(y) \rightarrow \eta^{\sigma \rho}$), and all the $k$-dependence becomes
\begin{equation}
\int \frac{d^4 k}{(2 \pi)^4} (k^2+m^2)^{-3} = \frac{i}{32 \pi^2 m^2}.
\end{equation} 

We obtain the following expression for $KE_{phys}$: 
\begin{equation}
\label{KEphys}
\begin{split}
&KE_{phys}(x) = \frac{1}{2} \eta^{\sigma \rho} \frac{i}{32 \pi^2 m^2} \bigg[ i ( \frac{1}{6} - \xi ) \partial_{\sigma}[R C_{\alpha}] \big( - \frac{1}{3} \delta^{\alpha}_{\,\,\, \rho} \big) \\
&+ \frac{i}{3} \partial_{\sigma}[R_{\alpha}^{\,\,\, \nu} C_{\beta}] \big(\frac{1}{3} \delta^{\alpha}_{\rho} \delta^{\beta}_{\nu} + \frac{1}{3} \delta^{\alpha}_{\nu} \delta^{\beta}_{\rho} - \frac{4}{3} \delta^{\beta}_{\rho} \delta^{\alpha}_{\nu} - \frac{4}{3} \delta^{\beta}_{\nu} \delta^{\alpha}_{\rho} - \frac{4}{3} \eta^{\beta \alpha} \eta_{\rho \nu} \big) \\
&+ \frac{i}{3} \partial_{\sigma}[R^{\mu \,\,\, \nu}_{\,\,\, \alpha \,\,\, \beta} C_{\gamma}] \bigg(-\frac{8}{5} \big[ \eta_{\rho \mu} \big( \eta^{\beta \gamma} \delta^{\alpha}_{\nu} + \eta^{\alpha \gamma} \delta^{\beta}_{\nu} + \eta^{\beta \alpha} \delta^{\gamma}_{\nu} \big) + \eta_{\rho \nu} \big( \eta^{\beta \gamma} \delta^{\alpha}_{\mu} + \eta^{\alpha \gamma} \delta^{\beta}_{\mu} + \eta^{\beta \alpha} \delta^{\gamma}_{\mu} \big) \\
&+ \delta^{\alpha}_{\rho} \big( \delta^{\beta}_{\mu} \delta^{\gamma}_{\nu} + \delta^{\beta}_{\nu} \delta^{\gamma}_{\mu} + \eta_{\mu \nu} \eta^{\beta \gamma} \big) + \delta^{\beta}_{\rho} \big( \delta^{\alpha}_{\mu} \delta^{\gamma}_{\nu} + \delta^{\alpha}_{\nu} \delta^{\gamma}_{\mu} + \eta_{\mu \nu} \eta^{\alpha \gamma} \big) + \delta^{\gamma}_{\rho} \big( \delta^{\beta}_{\mu} \delta^{\alpha}_{\nu} + \delta^{\beta}_{\nu} \delta^{\alpha}_{\mu} + \eta_{\mu \nu} \eta^{\beta \alpha} \big)  \big]  \\
&+2 \big[ \eta^{\alpha \beta} \big(\delta^{\gamma}_{\rho} \eta_{\mu \nu} + \delta^{\gamma}_{\nu} \eta_{\rho \mu} + \delta^{\gamma}_{\mu} \eta_{\rho \nu} \big) +  \eta^{\alpha \gamma} \big(\delta^{\beta}_{\rho} \eta_{\mu \nu} + \delta^{\beta}_{\nu} \eta_{\rho \mu} + \delta^{\beta}_{\mu} \eta_{\rho \nu} \big) +  \eta^{\gamma \beta} \big(\delta^{\alpha}_{\rho} \eta_{\mu \nu} + \delta^{\alpha}_{\nu} \eta_{\rho \mu} + \delta^{\alpha}_{\mu} \eta_{\rho \nu} \big) \big] \bigg) \\
&+ i \partial_{\sigma}[C_{\alpha} (\frac{1}{6}-\xi ) R] \big(-\frac{2}{3} \delta^{\alpha}_{\rho} \big) + i \partial_{\sigma} [I_{\alpha \beta \gamma}] \frac{20}{3} \big( \delta^{\beta}_{\rho} \eta^{\gamma \alpha} + \delta^{\alpha}_{\rho} \eta^{\gamma \beta} + \delta^{\gamma}_{\rho} \eta^{\beta \alpha} \big) \bigg] .
\end{split} 
\end{equation}
It is a function of $x$ since $y \rightarrow 0 $ was equivalent to $x' \rightarrow x$.

The expression is only valid for a non-zero $m$, and a very small mass is expected since dark energy acts on large scales.  With the help of xTensor in {\it Mathematica}, we then evaluate this expression for the 1st-order FLRW metric.  After a lengthy calculation, we arrive at a lengthy expression for $KE_{phys}(x)$ that depends on scale factor $a$, radial distance $r$ (since our FLRW perturbations depended on radial distance), $w$, mass $m$, non-minimal coupling $\xi$, and the constant of integration $S$ mentioned earlier, and we have kept all terms to first order in $S$.  The expression is far too long to show, but the relevant {\it Mathematica} files containing the expression are posted on Google Drive at 
https://goo.gl/5gY1mf.  

\bigskip
\begin{center}
{\bf Kinetic Energy Results}
\end{center}

We exhibit the behavior of $KE_{phys}$ in the figures in this section, varying different parameters involved in the expression for kinetic energy.  For all of these plots, we use $\rho_{DE0} = 4.12 \times 10^{-9}$ Mpc$^{-2}$, obtained from the best-fit values from Planck \cite{Planck}.  Most data sets of Type Ia supernovae data run from redshift $z \approx 0$ to $z \approx 2$ \cite{1401.4064, 1105.3470}, with the bulk of the constraining power coming from low redshift data.  For light traveling along a null geodesic in FLRW space dominated by dark energy with constant $w<-1$ but close to $-1$, a redshift of $z=2$ corresponds to $r=5410$ Mpc.  We will use this value of $r$ in our kinetic energy plots, but it turns out that the kinetic energy is almost completely independent of the radial distance $r$ at which we evaluate the kinetic energy, as we discuss below.  All the plots indicate that the renormalized kinetic energy of the dark energy field with a constant value of $w$ that is close enough to $-1$ and less than $-1$ is positive during dark energy domination.

\begin{figure}
\fbox{\includegraphics[scale=1.2]{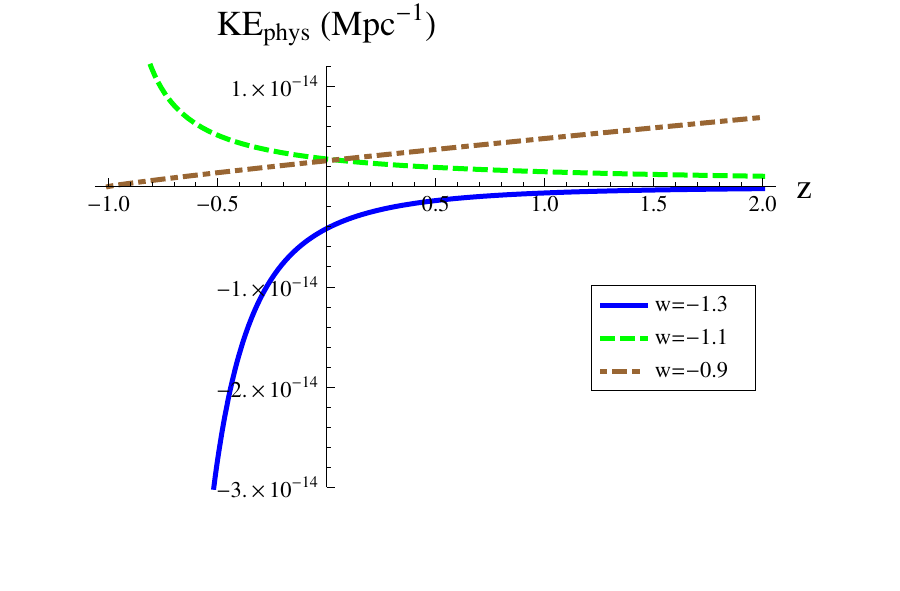}}
\caption{\small{$KE_{phys}$ vs. $z$ for $r=5410$ Mpc.  $KE_{phys}$ is plotted in units of Mpc$^{-1}$.  $m=3.2 \times 10^{-34}$ eV, $\xi =0$, and $S=10^{-5}$.  The range of $z$ is from $z \rightarrow -1$ (infinite future) to $z=2$. }}
\label{KEplot_z}
\end{figure}

Figure (\ref{KEplot_z}) shows positive kinetic energy plotted from $z=2$ to $z \rightarrow -1$ (the infinite future) for three choices of constant $w$.  Technically, because we have been considering dark energy as the only component present in the universe, our plot is only valid starting from the beginning of dark energy domination (roughly $z \sim 0.6$ and smaller redshift values).  The other free parameters are chosen to be reasonable: a very small mass value to correspond to a very long range for the dark energy field $\phi$, a small value for $\xi$ ($\xi=0$), and a small $S$ value ($|S| \sim 10^{-5}$ from our previous work in \cite{1507.06492}).  (See plot captions for specific parameter choices.) This plot is for $r=5410$ Mpc (corresponding to $z=2$ as discussed earlier), the upper bound of length scale covered by Type Ia supernovae data, but the plot is virtually unchanged for any large length scale, including all scales covered by the data (from $z=0$ to $z=2$).  From the observational data constraints quoted earlier in this paper, we know that $w$ cannot deviate too much from $-1$.  For a very negative value of $w$, such as $w=-1.3$, the kinetic energy is negative for all $z$ shown.  For $w= -1.1$ and $w= -0.9$, the kinetic energy is positive for all $z$ shown.  The $w=-1.1$ plot continues to increase into the future, and the plot for $w=-0.9$ approaches $0$ as $z \rightarrow -1$ (and this is true for non-phantom values, $-1/3<w<-1$).  Notice that the kinetic energy, as one might expect, does not change sign for all $z$ for any given value of $w$.  It is clear from Figure (\ref{KEplot_z}) that the value of $w$ for which the kinetic energy goes from negative to positive is between $-1.1$ and $-1.3$ when $\xi=0$ as in the figure.  We will return to this point later in our discussion.  

\begin{figure}
\fbox{\includegraphics[scale=1.2]{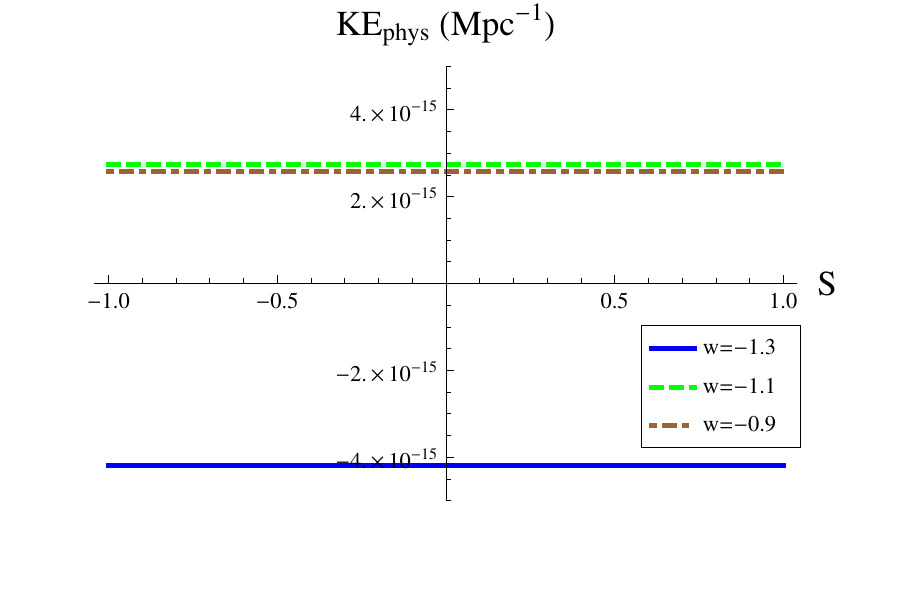}}
\caption{\small{$KE_{phys}$ vs. $S$ for $a=1$ (present day), $r=5410$ Mpc.  $KE_{phys}$ is plotted in units of Mpc$^{-1}$.  $m=3.2 \times 10^{-34}$ eV and $\xi =0$.  From matching with data, $S$ is expected be small, and $KE_{phys}$ is virtually unaffected by its size, not only for $a=1$ but virtually at any time.  So the contribution that is 0th order in $S$ in the kinetic energy dominates. }}
\label{KEplot_S}
\end{figure}

\begin{figure}
\fbox{\includegraphics[scale=1.2]{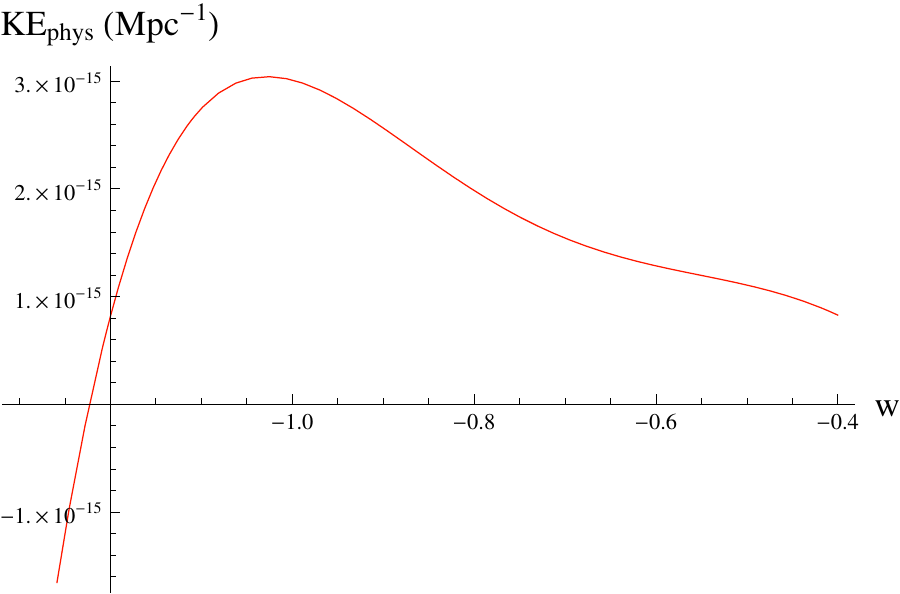}}
\caption{\small{$KE_{phys}$ vs. $w$ for $a=1$ (present day), $r=5410$ Mpc.  $KE_{phys}$ is plotted in units of Mpc$^{-1}$.  $m=3.2 \times 10^{-34}$ eV, $\xi =0$, and $S=10^{-5}$.  
The kinetic energy is negative for $w$ a little less than $-1.22$, and the same holds for any time since the beginning of dark energy domination. }}
\label{KEplot_w}
\end{figure}

\begin{figure}
\fbox{\includegraphics[scale=1.2]{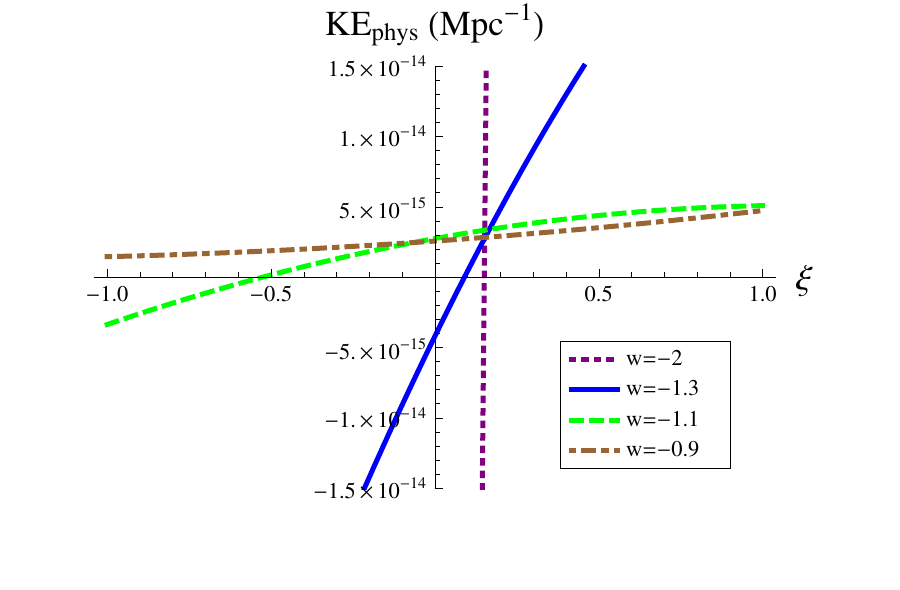}}
\caption{\small{$KE_{phys}$ vs. $\xi$ for $a=1$ (present day), $r=5410$ Mpc.  $KE_{phys}$ is plotted in units of Mpc$^{-1}$.  $m=3.2 \times 10^{-34}$ eV and $S=10^{-5}$. }}
\label{KEplot_xi}
\end{figure}

Figures (\ref{KEplot_S} - \ref{KEplot_xi}) have the same choices of reasonable parameter values (as listed in the figure captions).  The positivity or negativity of the kinetic energy is independent of $m$ since $KE_{phys}$ simply scales with $m$, as can be seen from Eq. (\ref{KEphys}).  Figure (\ref{KEplot_S}) is plotted for the present ($a=1$) and shows how little the 1st-order FLRW perturbation terms (which are all linearly proportional to $S$) affect the kinetic energy.  Clearly, the dominant contribution is from the terms that are 0th-order in $S$, and this is in fact true for any value of $a$ from the beginning of dark energy domination onward.  The value of $a$ tends to change simply the scale of the plot and not the distinct shape of it.  So the kinetic energy can therefore be positive even in 0th-order FLRW space (i.e., even when $S=0$). Because of this lack of dependence on the FLRW perturbations, the kinetic energy plots in Fig. (\ref{KEplot_z}) and the other figures therefore are similarly independent of the value of $r$ since $r$-dependence is only present in the 1st-order FLRW perturbations.  

Figure (\ref{KEplot_w}) shows positive kinetic energy for $\xi=0$ for a range of values for $w$ at present ($a=1$), with the kinetic energy being negative around $w \leq -1.22$ (and this is true for other any value of $a$ from the beginning of dark energy domination onward).  As can be seen from the previous figures and discussion, for a fixed value of $\xi$ and $w$, varying the other parameters does not affect the positivity of the kinetic energy.  The only parameters that noticeably affect the positivity of the kinetic energy is $\xi$ and $w$, and all the figures we have discussed so far (Figs. (\ref{KEplot_z} - \ref{KEplot_w})) have $\xi=0$.  Figure (\ref{KEplot_xi}) shows the kinetic energy over a range of values of $\xi$.  We see that kinetic energy is positive for all three aforementioned choices of $w$, along with an extreme choice of $w=-2$, for some values of $\xi$.  So with an appropriate choice of $\xi$, the kinetic energy for a very negative choice of $w=-1.3$ (and even $w=-2$) is positive for any value of $a$ from the beginning of dark energy domination onward, in contrast to the negative kinetic energy displayed in Fig. (\ref{KEplot_z}) for $w=-1.3$ when $\xi=0$.  

Regarding observational constraints on $\xi$, there are some phenomenological fifth-force constraints on the effective gravitational constant, $G_{eff}$, which is obtained from the terms involving the Ricci scalar in the action, and which depends on $\xi$:  
\begin{equation}
\label{G_eff}
\frac{R}{16 \pi G_{eff}} = \frac{R}{16 \pi G} - \frac{1}{2} \xi R \phi^2,
\end{equation}  
where $G$ is the canonical gravitational constant.  Since dark energy is effectively only active on cosmological scales, we will not consider small-scale constraints such as Solar System fifth-force constraints (which are overcome by screening in many models).  The ones that are relevant here in constraining $\xi$ should be large-scale constraints, and these are not as stringent.  For a low-mass scalar field such as $\phi$, galaxy mass component separation implies \cite{1807.01482}
\begin{equation}
\frac{G_{eff} - G}{G} \lesssim 10^{-4}.
\end{equation}
Another constraint arising from the measurement of the primordial abundance of deuterium in quasar absorption systems \cite{0311334, 1710.01120, 0503502} is
\begin{equation}
\frac{G_{BBN}}{G} = 1.01^{0.20}_{-0.16},
\end{equation}
where $G_{BBN}$ is the value during Big Bang Nucleosynthesis.  However, for our non-minimally coupled scalar, these observational constraints can be easily met with virtually no effect on $\xi$ by simply choosing the initial conditions of $\phi$ appropriately.  For late cosmological times, a constant-$w$ model of dark energy is supported by data, and for a constant-$w$ model during late times like we are dealing with in this work, Type Ia supernovae data do not constrain $\xi$ since $H$ can be written in terms of the energy density components without any dependence on $\phi$ or $\xi$.  
  
It is generally expected theoretically that $\xi$ is somewhat small, and from Figure (\ref{KEplot_xi}), we see that the most negative of $w$ values allowed from data constraints mentioned in the Introduction section can give positive kinetic energy for relatively small values of $\xi$, even for $w=-2$ and smaller, which is disfavored by observational constraints.  The important point here is that kinetic energy is positive for dark energy for all values of $w<-1$ allowed by data constraints quoted in the Introduction, even if $\xi=0$.


\bigskip
\begin{center}
{\bf Conclusion}
\end{center}

For standard cosmology, it is well-known that dark energy as a classical scalar field is ill-defined for $w<-1$ in 0th-order FLRW space, and as discussed in the Introduction section, the condition for positive kinetic energy for the dark energy scalar field in 1st-order FLRW implies that the perturbation $|\delta| > 1$, which is inconsistent with the perturbative assumption of 1st-order FLRW space \cite{1507.06492}.  

In this work, we have treated the scalar field as a quantum field with a small mass, obtained an approximate expression for the Green's function to 5th order (Eq. (\ref{G5})) using adiabatic expansion in Riemann normal coordinates, and calculated an expression for the kinetic energy that has been renormalized via adiabatic subtraction (and the expression is available in {\it Mathematica} files at the link quoted earlier in this work).  We find that the kinetic energy, somewhat surprisingly, is positive for constant $w<-1$.  For $\xi=0$, kinetic energy is positive for $w \gtrsim -1.22$, which includes virtually all values of constant $w$ allowed by cosmological data constraints, and more negative values of $w$ give positive kinetic energy for non-zero values of $\xi$.    
And we found this to be the case even at 0th-order FLRW (as illustrated by Figure (\ref{KEplot_S})).  We also confirmed that the dark energy field has positive kinetic energy for $w>-1$, as expected.  

This result gives credence and a more natural framework for observational data that suggest $w<-1$.  Without modifying gravity, flipping the sign in front of the kinetic term, or leaving the confines of general relativity, we have shown that a dark energy field with $w<-1$ is a viable option.  In principle, one could go further and keep more terms in the adiabatic expansion of the Green's function, and one could take into account effects due to the interaction potential $V(\phi)$ of the field in the calculation of the kinetic energy.  We expect these differences to be small, though, as discussed in the other sections.  Also, our results are generally applicable for a massive free field or a field with a small potential in a 0th- or 1st-order FLRW background dominated by a fluid with a constant $w$.

\bigskip
\begin{center}
{\bf Acknowledgements}
\end{center}

KJL is grateful for support from the LaGrange College Summer Research Grant Award.  



\end{document}